\documentclass[pra,preprint,superscriptaddress,amsmath,amssymb,showpacs]{revtex4}
\usepackage{bm}
\usepackage{amsmath}
\usepackage[dvips]{graphicx}
\begin{document}

\title{High-energy $e^+e^-$ photoproduction cross section close to the end of spectrum}

\author{A. Di Piazza}
\email{dipiazza@mpi-hd.mpg.de}
\affiliation{Max-Planck-Institut f\"ur Kernphysik, Saupfercheckweg 1, 69117 Heidelberg, Germany}

\author{A. I. Milstein}
\email{milstein@inp.nsk.su}
\affiliation{Max-Planck-Institut f\"ur Kernphysik, Saupfercheckweg 1, 69117 Heidelberg, Germany}
\affiliation{Budker Institute of Nuclear Physics, 630090 Novosibirsk, Russia}

\date{\today}

\begin{abstract}
We consider the cross section of electron-positron pair production by a high-energy photon in a strong Coulomb field close to the end of electron or positron spectrum. We show that the cross section essentially differs from the result obtained in the Born approximation as well as form the result which takes into account the Coulomb corrections under assumption that both electron and positron are ultrarelativistic. The cross section of bremsstrahlung in a strong Coulomb field by a high-energy electron is also obtained in the region where the final electron is not ultrarelativistic.
\end{abstract}

\maketitle

\section{Introduction}

The production of an electron-positron ($e^+e^-$) pair by a photon in a strong atomic field has been investigated already since many years both theoretically and experimentally because of the importance of this process for various applications, see Refs. \cite{HGO1980,Hubbell2000}. The cross section of this process in the Born approximation is known for arbitrary energy $\omega$ of the incoming photon, Refs. \cite{BH1934,Racah1934} (we set $\hbar=c=1$ throughout the paper). The effect of screening in this approximation  can be easily taken into account using the atomic form factor \cite{JLS1950}. For heavy atoms, it is however necessary to take into account the Coulomb corrections. These corrections are higher-order terms of the perturbation theory with respect to the parameter $Z\alpha$, where $Z$ is the atomic charge number and $\alpha=e^2\approx 1/137$ is the fine-structure constant, with $e$ being the absolute value of the electron charge. The formal expression of the Coulomb corrections, exact in $Z\alpha$ and $\omega$, was
derived in Ref. \cite{Overbo1968}. This expression has a very complicated form which leads to difficulties in numerical computations. The difficulties grow as $\omega$
increases, so that numerical results have been so far obtained only for $\omega<12.5\,$ MeV, Ref. \cite{SudSharma2006}.

In the high-energy region $\omega\gg m$ ($m$ is the electron mass), considerations become greatly simplified. As a result, a simple form of the Coulomb corrections was obtained in Refs. \cite{BM1954,DBM1954} in the leading approximation with respect to $m/\omega$. However, this result has good accuracy only at energies $\omega \gtrsim 100\,$MeV. The theoretical description of the Coulomb corrections for the total cross section at intermediate photon energies ($5 \div 100\,$ MeV) was based during a long time on the ``bridging'' expression derived in \cite{Overbo1977}. This expression is actually an extrapolation of the results obtained at $\omega<5\,$MeV. Results for the spectrum of one of the created particles at intermediate $\omega$ were practically absent. Recently, an important step was made in Ref. \cite{LMS2004} where the first corrections of the order of $m/\omega$ to the spectrum as well as to the total cross section of $e^+e^-$ photoproduction in a strong atomic field were derived. The correction to spectrum was obtained in the region where both produced particles are relativistic. It turns out that this correction is antisymmetric with respect to replacement $\epsilon_+\leftrightarrow\epsilon_-$, where $\epsilon_+$ and $\epsilon_-$ are the energy of the positron and the electron, respectively. Since the correction to the total cross section resulted to be very large, it is not related to the central region, where the created electron and positron are ultrarelativistic, and it comes from the region close to the end of spectrum where  $\epsilon_+\sim m$ or $\epsilon_-\sim m$. In Ref. \cite{LMS2004}, the correction to the total cross section was obtained with the use of the dispersion relation for the forward Delbr\"uck scattering amplitude but not by the direct integration of the spectrum. Note that the account for the correction to the total cross section leads to good  description of available experimental data at intermediate photon energies, Refs. \cite{LMS2003}.

In the present paper, we calculate the electron (positron) spectrum in the process of $e^+e^-$ photoproduction in a strong Coulomb field in the case $\epsilon_-\sim m$ ($\epsilon_+\sim m$) and $\omega\gg m$. We show that the Coulomb corrections drastically differ form that obtained in the region where $\epsilon_+\gg m$ and $\epsilon_-\gg m$. In an analogous way, we have also derived the spectrum of bremsstrahlung in a strong Coulomb field in the region where the radiated photon has the energy close to that of the initial electron.

\section{General discussion}
Following the usual Feynman rules (see, e. g., Ref. \cite{BLP1980}), the $e^+e^-$ photoproduction cross section, at leading order in the interaction between the photon field and the electron-positron field, averaged over the polarization of the incoming photon and summed up over polarizations of electron and positron has the form
\begin{eqnarray}\label{sigmaee0}
d\sigma&=&\frac{4\pi\alpha}{2\omega}\left(-\frac{1}{2}\right)2\pi\delta(\omega-\epsilon_+-\epsilon_-)\frac{d\bm p_+}{(2\pi)^3}\frac{d\bm p_-}{(2\pi)^3}\sum_{\lambda_-,\,\lambda_+}{\cal M}^{\mu}{\cal M}^*_{\mu}
\,,\nonumber\\
{\cal M}^{\mu}&=&
\int d\bm x\bar U_{\bm p_-,\lambda_-}(\bm x)\gamma^\mu V_{\bm p_+,\lambda_+}(\bm x)\mbox{e}^{i\bm k\cdot\bm x}\,,
\end{eqnarray}
where $(\omega,\bm k)$ is the four-momentum of the photon, $(\epsilon_-,\bm p_-)$ and $(\epsilon_+,\bm p_{+})$ are the four-momenta of the electron and the positron, respectively, and $\lambda_{-}$ and $\lambda_+$ are their polarization indexes. Also, $U_{\bm p_-,\lambda_-}$ and $V_{\bm p_+,\lambda_+}$ are the corresponding positive-energy and negative-energy wave functions in a strong Coulomb field and $\gamma^\mu$ are the Dirac matrices. In the following, we will calculate the spectrum, i.e., the cross section integrated over the angles of the vectors $\bm p_+$ and $\bm p_-$. Due to rotational symmetry, this quantity is independent of the direction of the photon momentum $\bm k$. Therefore, we can average it over this direction, i.e., integrate both sides of Eq. (\ref{sigmaee0}) over $d\Omega_{\bm k}/(4\pi)$. This results in the replacement
\begin{equation}
\mbox{e}^{i\bm k\cdot(\bm x-\bm y)}\longrightarrow\frac{\sin(\omega R)}{\omega R}   \,,
\end{equation}
where $R=|\bm y-\bm x|$. We can also use the known relations \cite{MS1983}
\begin{eqnarray}\label{relg}
&&\sum_\lambda \int \frac{d\Omega_{\bm p}}{(2\pi)^3}U_{\bm p,\lambda}(\bm x)\bar U_{\bm p,\lambda}(\bm y)
=\frac{i}{2\pi p\epsilon}\delta G(\bm x,\bm y\,|\epsilon)\,,\nonumber\\
&&\sum_\lambda \int \frac{d\Omega_{\bm p}}{(2\pi)^3}V_{\bm p,\lambda}(\bm x)\bar V_{\bm p,\lambda}(\bm y)
=\frac{-i}{2\pi p\epsilon}\delta G(\bm x,\bm y\,|-\epsilon)\,,
\end{eqnarray}
where $\epsilon>0$ and $\delta G(\bm x,\bm y\,|\pm\epsilon)$ is the discontinuity on the cut of the electron Green's function in the Coulomb field, to obtain
\begin{eqnarray}\label{sigmaeeGG}
\frac{d\sigma}{d\epsilon_-}&=&-\frac{\alpha}{2\omega }
\int\!\!\int d\bm x\, d\bm y\,  \mbox{Sp}\left[\gamma^\rho\delta G(\bm x,\bm y\,|\epsilon_- -\omega)\gamma_\rho
 \delta G(\bm y,\bm x\,|\epsilon_-)\right]\frac{\sin(\omega R)}{\omega R}\,,
\end{eqnarray}
Since the representation (\ref{relg}) of the quantity $\delta G(\bm y,\bm x\,|\epsilon_-)$ is independent of the basis employed, we can also write $\delta G(\bm y,\bm x\,|\epsilon_-)$ in the form
\begin{eqnarray}\label{relg1}
&&\delta G(\bm y,\bm x\,|\epsilon_-)=-\frac{i}{\beta_-}\sum_{j,\sigma,\mu}
 U_{j,\sigma,\mu}(p_-,\bm y)\bar U_{j,\sigma,\mu}(p_-,\bm x)\, ,
\end{eqnarray}
and represent the spectrum as follows
\begin{eqnarray}\label{sigmaee1}
\frac{d\sigma}{d\epsilon_-}&=&\frac{i\alpha}{2\omega \beta_-}\sum_{j,\sigma,\mu}
\int\!\!\int d\bm x\, d\bm y \bar U_{j,\sigma,\mu}(p_-,\bm x)\gamma^\rho\delta G(\bm x,\bm y\,|\epsilon_--\omega)\gamma_\rho
 U_{j,\sigma,\mu}(p_-,\bm y)\frac{\sin(\omega R)}{\omega R}\,,
\end{eqnarray}
where  $\beta_-=p_-/\epsilon_-$,  $\sigma=\pm 1$. Here, $U_{j,\sigma,\mu}(p,\bm x)$ is the positive-energy wave function with total angular momentum $j$, parity $(-1)^{j+\sigma/2}$, and projection $\mu$ of the total angular momentum along some quantization axis. The explicit form of this function is presented in Appendix \ref{A}.

Below, we assume that $\omega\gg m$. If the momentum of the electron is $p_-\sim m$, then the formation length of the process is of the order of the Compton wavelength $1/m$ and the positron is ultrarelativistic with a typical angular momentum $l_+\sim \omega/m\gg1$. This circumstance allows us to use the quasiclassical Green's function obtained in Refs. \cite{MS1983a,MS1983} starting from a convenient integral representation derived in Ref. \cite{MS1982} of the exact Green's function of the Dirac equation in a Coulomb field. For the reader's convenience, we present the formula for the discontinuity of this Green's function in Appendix \ref{A}, Eqs. (\ref{dgf}). Moreover, it can be seen that if $p_-\ll p_+$, then the main contribution to the integral in Eq. (\ref{sigmaee1}) is given by the region where the angles between vectors $\bm x$ and $\bm y$ are not small (or close to $\pi$). This is the region which also gives the main contribution to the bound-free photoproduction cross section at $\omega\gg m$, see Ref. \cite{MS1993}. In this region the expression in Eq. (\ref{dgf}) becomes essentially simpler, see Eq. (\ref{dgf1}). By substituting Eq. (\ref{dgf1}) in Eq. (\ref{sigmaee1}) and by keeping the terms which do not contain highly oscillating functions, we arrive at
\begin{eqnarray}\label{sigmaee2}
\frac{d\sigma}{d\epsilon_-}&=&\frac{\alpha}{4\pi\omega \beta_-}\sum_{j,\sigma,\mu}
\int\!\!\int \frac{d\bm x\, d\bm y}{R^2} \bar U_{j,\sigma,\mu}(p_-,\bm x)(\gamma^0\cos\phi-i\bm\gamma\cdot\bm n\sin\phi)
 U_{j,\sigma,\mu}(p_-,\bm y)\,,
\end{eqnarray}
where $\phi=\epsilon_- R+2Z\alpha s$ (the notation is explained in Eq. (\ref{dgf1})). By employing the explicit form (\ref{wf}) of the electron wave function, we finally arrive at the following expression of the cross section of pair production in the case of slow electron
\begin{eqnarray}\label{sigmaee3}
\frac{d\sigma}{d\epsilon_-}&=&\frac{\alpha}{4\pi^2\omega \beta_-}\sum_{j,\sigma} \left(j+\frac{1}{2}\right)
\int\!\!\int \frac{d\bm x\, d\bm y}{x y R^2}(F\cos\phi-T\sin\phi)\, ,\nonumber\\
F&=&f(x)f(y)P_l(t)+g(x)g(y)P_{l'}(t)\,,\nonumber\\
T&=&\frac{1}{R}\{f(x)g(y)[xP_{l'}(t)-yP_l(t)]+g(x)f(y)[yP_{l'}(t)-xP_l(t)]\}\,,
\end{eqnarray}
where $t=\bm x\cdot\bm y/(xy)$, $l=j+\sigma/2$, $l'=j-\sigma/2$, and $P_l(t)$ are the Legendre polynomials. The cross section on the other end of the spectrum, i.e. in the case of slow positron with momentum $p_+\sim m$, is given by Eq. (\ref{sigmaee3}) with the replacement $\epsilon_-\to \epsilon_+$, $\beta_-\to\beta_+=p_+/\epsilon_+$, and $Z\to -Z$. With the same procedure, one can also derive the spectrum of  bremsstrahlung by a high-energy electron for the case where the final electron with momentum $p_1$ and energy $\epsilon_1$ is slow ($p_1\sim m$). It turns out that this spectrum is given by the same formula (\ref{sigmaee3}) with the obvious substitutions $p_-\to p_1$ and $\epsilon_-\to\epsilon_1$. Note that the correction to the bremsstrahlung spectrum of the order of $m/\epsilon_1$ in the case of initial and final electrons both ultrarelativistic was obtained recently in Ref. \cite{LMS2004} from the corresponding results for pair production, and in Ref. \cite{LMSSch2005} directly from the matrix element of bremsstrahlung.

It can be shown that three integrations in Eq. (\ref{sigmaee3}) can be performed analytically and the spectrum $d\sigma/d\epsilon_-$ becomes
\begin{eqnarray}\label{sigmaee4}
\frac{d\sigma}{d\epsilon_-}&=&\frac{2\alpha}{\omega \beta_-}\sum_{j,\sigma}  \left(j+\frac{1}{2}\right)
\int\limits_{0}^{\infty} \!\!\int\limits_{0}^{\infty} \,dx\, dy\int\limits_{-1}^{1}dt\,\frac{xy}{ R^2}(F\cos\phi-T\sin\phi)\,.
\end{eqnarray}
It is convenient to multiply the integrand in this formula by unity written in the form
\begin{equation}
 1\equiv \int\limits_0^1 du\, 2u\delta\left(u^2-\frac{R^2}{(x+y)^2}\right) \,,
\end{equation}
to change the order of integration over the variables $t$ and $u$, and to take the integral over $t$ (by exploiting the $\delta$-function). After that, we pass from the variables $x$ and $y$ to the variables $\rho$ and $v$ such that $x=\rho (1+v)/2$ and $y=\rho (1-v)/2$. As a result we obtain
\begin{eqnarray}\label{sigmaeefinal}
\frac{d\sigma}{d\epsilon_-}&=&\frac{\alpha}{\omega \beta_-}\sum_{j,\sigma}  \left(j+\frac{1}{2}\right)
\int\limits_0^\infty\rho\, d\rho\int\limits_{-1}^1 dv \int\limits_{|v|}^1 \frac{du }{u}(F\cos\Phi-T\sin\Phi)\, ,\nonumber\\
F&=&f(x)f(y)P_l(t_0)+g(x)g(y)P_{l'}(t_0)\,,\nonumber\\
T&=&\frac{1}{2u}\Big\{f(x)g(y)[(1+v)P_{l'}(t_0)-(1-v)P_l(t_0)]\nonumber\\
&&+g(x)f(y)[(1-v)P_{l'}(t_0)-(1+v)P_l(t_0)]\Big\}\,,\nonumber\\
&&x=\rho \frac{1+v}{2}\,,\quad y=\rho \frac{1-v}{2}\,,\quad t_0= 2\frac{1-u^2}{1-v^2}-1 \,,\nonumber\\
&&\Phi= \epsilon_-u\rho+Z\alpha\ln\left(\frac{1+u}{1-u}\right)\,.
\end{eqnarray}
This formula  is still not convenient for numerical calculations because  of the strong oscillations of the integrand in the vicinity of the point $u=0$. In order to overcome this difficulty, we  write
\begin{equation}
F\cos\Phi-T\sin\Phi=\mbox{Re}M(u,v,\rho)\, , \quad M(u,v,\rho)=(F+iT)\mbox{e}^{i\Phi}\,.
\end{equation}
Then, by using the properties of the integrand $M(u,v,\rho)$, we make the following transformation
\begin{eqnarray}\label{transform}
&& \int\limits_{-1}^1 dv \int\limits_{|v|}^1 \frac{du }{u}\int\limits_0^\infty\rho\, d\rho\,M(u,v,\rho)=
\frac{1 }{2}\int\limits_{-1}^1 du\int\limits_{-1}^1 dv\int\limits_0^\infty\rho\, d\rho\,M(u,vu,\rho)\nonumber\\
&& =\frac{i }{2}
\int\limits_{0}^\infty dr \!\int\limits_{-1}^1 dv \!\int\limits_0^\infty\rho\, d\rho\,
\Big[M\Big(-1+ir,v (-1+ir),\rho\Big)-M\Big(1+ir,v (1+ir),\rho\Big)\Big].
\end{eqnarray}
In the first step we changed the integration order of the variables $u$ and $v$ and then we performed the change of variable $v\to vu$. In the second step, we changed the contour of integration, by exploiting the fact that the contribution along the straight path from the point $u_1=1+i\infty$ to the point $u_2=-1+i\infty$ vanishes due to the exponential function $\exp(i\Phi)$ (see Eq. (\ref{sigmaeefinal}) and also Fig. 1). This form of integral is appropriate for numerical calculations. Note that the integral (\ref{transform}) has zero imaginary part, so that it is not necessary to take real part of it afterwards.

\section{Asymptotics $\beta_-\to 0$}
Let us consider the spectrum Eq. (\ref{sigmaeefinal}) in the limit $\beta_-=p_-/\epsilon_-\to 0$. Substituting the asymptotics (\ref{leas}) of the functions $f(r)$ and $g(r)$ in Eq. (\ref{sigmaeefinal}), we obtain
\begin{eqnarray}\label{sigmaeefinalA}
\frac{d\sigma}{d\epsilon_-}&=&\frac{\alpha\pi}{\omega m^2 (Z\alpha)}\sum_{j} \left(j+\frac{1}{2}\right)
\int\limits_0^\infty\rho\, d\rho\int\limits_{-1}^1 dv \int\limits_{|v|}^1 \frac{du }{u}(F_0\cos\Phi_0-T_0\sin\Phi_0)\, ,\nonumber\\
F_0&=&\{[\kappa^2+(Z\alpha)^2]J_{2\gamma}(\eta_1)J_{2\gamma}(\eta_2)+\frac{\eta_1\eta_2}{4}J_{2\gamma}'(\eta_1)J_{2\gamma}'(\eta_2)\}
[P_{j+1/2}(t_0)+P_{j-1/2}(t_0)]\nonumber\\
&&-\frac{1}{2}\left(j+\frac{1}{2}\right)[J_{2\gamma}(\eta_1)\eta_2J_{2\gamma}'(\eta_2)+J_{2\gamma}(\eta_2)\eta_1J_{2\gamma}'(\eta_1)]
[P_{j+1/2}(t_0)-P_{j-1/2}(t_0)]\,,\nonumber\\
T_0&=&\frac{Z\alpha}{2u}\Big\{v[J_{2\gamma}(\eta_1)\eta_2J_{2\gamma}'(\eta_2)-J_{2\gamma}(\eta_2)\eta_1J_{2\gamma}'(\eta_1)]
[P_{j+1/2}(t_0)+P_{j-1/2}(t_0)]\nonumber\\
&&-4\left(j+\frac{1}{2}\right)J_{2\gamma}(\eta_1)J_{2\gamma}(\eta_2)[P_{j+1/2}(t_0)-P_{j-1/2}(t_0)]\Big\}\,,
\nonumber\\
&&\eta_1=2\sqrt{Z\alpha\rho (1+v)}\,,\quad \eta_2=2\sqrt{Z\alpha\rho(1-v)}\,,\nonumber\\
&&\Phi_0=u\rho+Z\alpha\ln\left(\frac{1+u}{1-u}\right)\,,\quad t_0= 2\frac{1-u^2}{1-v^2}-1 \,.
\end{eqnarray}
The integration over the variable $\rho$ can be taken by using the relation \cite{Grad}
\begin{eqnarray}\label{int}
&&\int\limits_0^\infty  dx \mbox{e}^{icx}J_\mu(a\sqrt{x})J_\mu(b\sqrt{x})=
\frac{i}{c}J_\mu\left(\frac{ab}{2c}\right)\exp\left(i\frac{\pi\mu}{2}-i\frac{a^2+b^2}{4c}\right)
\,,
\end{eqnarray}
The remaining integrations over the variables $v$ and $u$ can be performed numerically by employing the transformation (\ref{transform}). The largest contribution to the sum over $j$ is given by the term with $j=1/2$. The contribution of the term with $j=3/2$ is essentially smaller, while that with $j=5/2$ is less than one percent even for large $Z$. In our numerical calculations here we have included terms with $j=1/2$, $j=3/2$ and $j=5/2$. The results for $\omega d\sigma/d\epsilon_-$ in units of $\tilde{\sigma}=\alpha (Z\alpha)^3/m^2$ at zero electron velocity is shown in Fig. \ref{figV0} as a solid curve. At $Z\alpha\to 0$ we obtain $\omega d\sigma/d\epsilon_-=4\pi\tilde{\sigma}$ in agreement with previous results. In Ref. \cite{Deck1969} by Deck et al., the following formula for $\omega d\sigma/d\epsilon_-$ at zero electron velocity was suggested
\begin{eqnarray}\label{Deck}
\omega \frac{d\sigma^{(D)}}{d\epsilon_-}=4\pi\frac{\alpha (Z\alpha)^3}{m^2}\,\frac{2\pi Z\alpha}{\exp(2\pi Z\alpha)-1}\left(1-\frac{4\pi}{15} Z\alpha\right)\,.
\end{eqnarray}
This formula is shown in Fig. \ref{figV0} as a dashed curve and it is clear from the figure that Eq. (\ref{Deck}) is applicable only at small values of $Z\alpha$. Also, note that the cross section $\omega d\sigma_B/d\epsilon_-$ in the Born approximation vanishes in the limit $\beta_-\to 0$, since at $\beta_-\ll 1$ it scales as $\omega d\sigma_B/d\epsilon_-\approx 2\alpha(Z\alpha)^2\beta_-/m^2$. In Ref. \cite{SudSharma2006} the results for $\omega d\sigma/d\epsilon_-$ was obtained at $\omega=40\,$MeV and $\epsilon_-=1.008\,m$ (which corresponds to $\beta_-=0.1265$). These results are shown in Fig. \ref{figV0} as a dotted curve starting from $Z=11$. One sees an excellent agreement of our results with those of Ref. \cite{SudSharma2006}. At $Z=1$ there is disagreement because at small $Z$ and $\beta_-=0.1265$ the contribution of the Born term is also important. Finally, we observe that, as expected, the spectrum of positron at small positron velocity tends to zero because in this case the wave functions are exponentially small (see Appendix \ref{A}).

\section{Cross section at non-zero electron velocity}
In order to obtain the spectrum for non-zero electron velocity, we substitute the explicit form of wave function (\ref{wf}) in Eq. (\ref{sigmaeefinal}) and use the relation
$F(\alpha,\beta_-,x)=\mbox{e}^xF(\beta_--\alpha,\beta_-,-x)$ for the confluent hypergeometric function. We come to the following expression
for the cross section at $p_-\sim m$
\begin{eqnarray}\label{sigmaeefinal1}
&&\frac{d\sigma}{d\epsilon_-}=\frac{\alpha}{4\omega \beta_- p_-^2}\sum_{L=1}^\infty L \mbox{e}^{\pi\nu}
\frac{|\Gamma(\gamma+1+i\nu)|^2}{[\Gamma(2\gamma+1)]^2}
\mbox{Re}\int\limits_{-1}^1 du\int\limits_{-1}^1 dv \int\limits_0^\infty d\rho\, \rho^{2\gamma+1}
(1-v^2u^2)^{\gamma}\mbox{e}^{i\Phi}{\cal M}\, ,\nonumber\\
&&{\cal M}=\frac{F_1F_2}{\gamma-i\nu}\left[i\nu \frac{m^2}{\epsilon_-^2}\Delta_+-L\left(1+\frac{\beta_-}{u}\right)\Delta_-\right]
-\frac{{\tilde F}_1{\tilde F}_2}{\gamma+i\nu}\left[i\nu \frac{m^2}{\epsilon_-^2}\Delta_++L\left(1-\frac{\beta_-}{u}\right)\Delta_-\right]
\nonumber\\
&&+\left[F_1{\tilde F}_2 (1+\beta_- v)+{\tilde F}_1F_2(1-\beta_- v)\right]\Delta_+\,,\nonumber\\
&&F_{1,2}=F(\gamma-i\nu,2\gamma+1,-i\rho(1\pm vu))\,,\quad  {\tilde F}_{1,2}=F(\gamma+1-i\nu,2\gamma+1,-i\rho(1\pm vu))\,, \nonumber\\
&&\Delta_{\pm}=P_{L}(\tilde{t})\pm P_{L-1}(\tilde{t})\,,\;\; \Phi=\left(\frac{u}{\beta_-}+1\right)\rho+Z\alpha\ln\left(\frac{1+u}{1-u}\right)\,,\;\; \tilde{t}=2\frac{1-u^2}{1-v^2 u^2}-1\,,
\end{eqnarray}
where $\nu=Z\alpha/\beta_-$ and $\gamma=\sqrt{L^2-(Z\alpha)^2}$. Then, we perform the transformation (\ref{transform}) and take analytically
the integral over the variable $\rho$ using the relation (see mathematical Appendix f in Ref. \cite{LLQM})
\begin{eqnarray}\label{rel1}
&&\int\limits_0^{\infty}\mbox{e}^{-\lambda z} z^{\gamma-1}F(\alpha,\gamma,kz) F(\alpha',\gamma,k'z)\,dz\nonumber\\
&&=\Gamma(\gamma)\lambda^{\alpha+\alpha'-\gamma}(\lambda-k)^{-\alpha} (\lambda-k')^{-\alpha'}
 F\left(\alpha,\alpha',\gamma,\frac{kk'}{(\lambda-k)(\lambda-k')}\right)\, ,
\end{eqnarray}
where $F(a,b,c,x)$ is the hypergeometric function. After that we take numerically the integrals over the variables $v$ and $u$.

The cross section in Eq. (\ref{sigmaeefinal1}) can be represented as
\begin{equation}\label{S}
\frac{d\sigma}{d\epsilon_-}=\sum_{L=1}^\infty \frac{d\sigma_L}{d\epsilon_-}\, .
\end{equation}
Unfortunately, the convergence of the series  (\ref{S}) is not fast, and it is necessary to take into account terms $d\sigma_L/d\epsilon_-$ with rather large $L$, especially at $\epsilon_-\gg m$. In order to overcome this problem, we make the following approximation
\begin{equation}\label{S1}
\sum_{L=1}^\infty\frac{d\sigma_L}{d\epsilon_-}\approx\sum_{L=1}^{L_0}\left(\frac{d\sigma_L}{d\epsilon_-}-\frac{d\sigma^A_L}{d\epsilon_-}\right)+\Sigma^A\, ,\quad  \Sigma^A =\sum_{L=1}^\infty \frac{d\sigma^A_L}{d\epsilon_-}\, ,
\end{equation}
where $L_0$ is a large integer and $d\sigma^A_L/d\epsilon_-$ is given by the expression for $d\sigma_L/d\epsilon_-$ with the replacement $\gamma=\sqrt{L^2-(Z\alpha)^2}\to L$. The convergence of the series $\sum_{L=1}^{\infty}(d\sigma_L/d\epsilon_--d\sigma^A_L/d\epsilon_-)$ is much faster than the convergence of $\sum_{L=1}^{\infty}d\sigma_L/d\epsilon_-$, and it is not necessary to take very large value of $L_0$ (we have seen that by choosing $L_0=5$, an accuracy of about $10\;\%$ is reached), while the series $\Sigma^A$ can be summed analytically. Since the summation is not straightforward, we report some steps in the next paragraph.

\subsection{Calculation of $\Sigma^A$}
It is convenient to perform calculation of $\Sigma^A$ starting from Eq. (\ref{sigmaeeGG}). After the replacement $\gamma\to L$, the expression for the discontinuity of the  electron Green's function is given by Eq. (\ref{dgf}). By using this expression and by passing to the same variables as in the derivation of Eq. (\ref{sigmaeefinal}), we can again employ the convenient transformation (\ref{transform}). In this way, the integral over the variable $v$ can be easily performed and one obtains
\begin{eqnarray}\label{SA}
&& \Sigma^A=-\frac{i\alpha}{4\omega p_-^2}\int\limits_{0}^\infty du \Big[ M^A(-1+iu)- M^A(1+iu)\Big]\, ,\nonumber\\
&&M^A(u)=\int\limits_0^\infty d\rho\,\rho^3\,\mbox{e}^{i\tilde{\Phi}}
\int\limits_{-\infty}^{+\infty}\frac{d\tau}{\sinh^2\tau}
\exp\left[i(2\nu\tau +\rho\coth\tau)\right]\nonumber\\
&&\times\bigg[\frac{1}{\beta_-}\left(1-\frac{u^2}{3}\right)J_0(w)-2iZ\alpha(1-u^2)\coth\tau\frac{J_1(w)}{w}\nonumber\\
&&+ \frac{2u}{3}\coth\tau J_0(w) -i\frac{\rho u (1-u^2)}{3\sinh^2\tau}\frac{J_1(w)}{w}\bigg]\,,\nonumber\\
&& w=\frac{\rho\sqrt{1-u^2}}{\sinh\tau}\,,\quad \tilde{\Phi}=\frac{\rho u}{\beta_-}+Z\alpha\ln\left(\frac{1+u}{1-u}\right)\,.
\end{eqnarray}
Since the contour of integration over $\tau$ passes in the positive direction around the point $\tau=0$, we can make a shift
$\tau\to \tau-i\pi/2$ and take the integral over $\tau$ in $M^A(u)$ using the relation \cite{Eerdelyi}
\begin{eqnarray}\label{rel2}
&&\int\limits_{-\infty}^{+\infty}\frac{d\tau}{\cosh\tau}\exp\left(2\lambda\tau -b\tanh\tau\right)
J_{2\mu}\left(\frac{a}{\cosh\tau}\right)=\mbox{e}^{-b} a^{2\mu} \frac{\Gamma(\frac{1}{2}+\lambda+\mu)\Gamma(\frac{1}{2}-\lambda+\mu)}{[\Gamma(2\mu+1)]^2}\nonumber\\
&&\times
F\left(\mu-\lambda+\frac{1}{2},\,2\mu+1,b+\sqrt{b^2-a^2}\right) F\left(\mu-\lambda+\frac{1}{2},\,2\mu+1,b-\sqrt{b^2-a^2}\right)\,.
\end{eqnarray}
Then we take the integral over $\rho$ with the help of Eq. (\ref{rel1}). Finally, we have
\begin{eqnarray}\label{MU}
&& M^A(u)=2i\left[-\frac{\nu}{\beta_-}\left(1-\frac{u^2}{3}\right)+Z\alpha(1-u^2)-\frac{2}{3}u\nu\right]I^{(2)}_{11}\nonumber\\
&&+(1-i\nu)\bigg\{\frac{2i}{u}(1-i\nu)\left(1-\frac{u^2}{3}\right)I^{(1)}_{22}- \frac{i}{u}(2-i\nu)\left(1-\frac{u^2}{3}\right)\left(I^{(1)}_{31}+I^{(1)}_{13}\right)  \,\nonumber\\
&&+\frac{2i}{3}(2-i\nu)\left(I^{(1)}_{31}-I^{(1)}_{13}\right) +\frac{i}{u}\left(1-\frac{u^2}{3}\right)\left(I^{(1)}_{12}+I^{(1)}_{21}\right)-\bigg[\frac{1}{\beta_-}\left(1-\frac{u^2}{3}\right)+\frac{2u}{3}\bigg]\left(I^{(2)}_{12}+I^{(2)}_{21}\right)\nonumber\\
&& +\bigg[\frac{1}{\beta_- u}\Big(1-\frac{u^2}{3}\bigg)+ \frac{2}{3}\left(2-u^2\right)\bigg]\left(I^{(2)}_{21}-I^{(2)}_{12}\right)\nonumber\\
&&+\bigg[\frac{2Z\alpha}{u}(1-u^2)
+\frac{i}{3}\bigg(4i\nu +\frac{3}{u^2}-5\bigg)\bigg]\left(I^{(1)}_{21}-I^{(1)}_{12}\right)\bigg\}\,,\nonumber\\
&&I^{(n)}_{jk}=(-1)^n\left.\frac{d^n I_{jk}(\lambda)}{d\lambda^n}\right\vert_{\lambda=-i(1+u/\beta_-)}\,,\nonumber\\
&&I_{jk}(\lambda)=\lambda^{j+k-2i\nu}(\lambda+i(1+u))^{i\nu-j}(\lambda+i(1-u))^{i\nu-k}\nonumber\\
&&\times F\left(j-i\nu,k-i\nu,2,\frac{1-u^2}{(\lambda+i(1-u))(\lambda+i(1+u))}\right)\,.
\end{eqnarray}
This expression is particularly suitable for numerical integration. In the next section we report our results obtained starting from Eqs. (\ref{sigmaeefinal1}) in the approximation (\ref{S1}).

\section{Results and discussion}
The cross section $d\sigma_B/d\epsilon_-$ of the pair production process in the Born approximation is well known (see, for example, Ref. \cite{BLP1980}). In our limit ($\omega\gg m$ and $p_-\ll \omega$), it has the form
\begin{equation}\label{born}
\frac{d\sigma_B}{d\epsilon_-}=\frac{\sigma_0}{\omega}\frac{2\epsilon_-}{p_-^3}\left[2\epsilon_- p_- \ln\left(\frac{\epsilon_-+p_-}{m}\right)
-p_-^2-m^2\ln^2\left(\frac{\epsilon_-+p_-}{m}\right)\right]\,,\quad \sigma_0=\frac{\alpha(Z\alpha)^2}{m^2}\,.
\end{equation}
In particular, at $p_-\ll m$ it is $d\sigma_B/d\epsilon_-=2\sigma_0p_-/m$, while at $p_-\gg m$ the same cross section has the asymptotics
\begin{equation}\label{born1}
\frac{d\sigma_B}{d\epsilon_-}=\frac{4\sigma_0}{\omega}\left[\ln\left(\frac{2\epsilon_-}{m}\right)-\frac{1}{2}\right]\,.
\end{equation}
Now, the leading Coulomb correction to $d\sigma/d\epsilon_-$ at $m\ll \epsilon_-\ll\omega$ reads \cite{DBM1954}
\begin{equation}\label{cc0}
\frac{d\sigma_C^{(0)}}{d\epsilon_-}=-\frac{4\sigma_0}{\omega}\,f(Z\alpha)\,,\quad f(Z\alpha)= \mbox{Re}[\psi(1+iZ\alpha)+C]\,,
\end{equation}
where $\psi(x)=d\ln \Gamma(x)/dx$, $C=0.577...$ is the Euler constant. The correction (\ref{cc0}) is independent of $\epsilon_-$ and it is the same for electron and positron (i. e., it is an even function of $Z\alpha$). The next-to-leading correction was calculated recently in Ref. \cite{LMS2004} and at $m\ll \epsilon_-\ll\omega$ it has the form
\begin{equation}\label{cc1}
\frac{d\sigma_C^{(1)}}{d\epsilon_-}=\frac{\sigma_0}{\omega}\,\frac{\pi^3 m}{2\epsilon_-} \mbox{Re}\, g(Z\alpha)\,,\quad
g(Z\alpha)=Z\alpha\frac{\Gamma(1-iZ\alpha)\Gamma(1/2+iZ\alpha)}{\Gamma(1+iZ\alpha)\Gamma(1/2-iZ\alpha)}\,.
\end{equation}
This correction has opposite sign for electron and positron since $g(x)$ is an odd function of $Z\alpha$ and it increases the cross section for slow electron while it decreases it for slow positron. In Fig. 3 and Fig. 4 we show our results for the Coulomb corrections $\omega \sigma_0^{-1}d\sigma_C/d\epsilon_{\mp}=\omega \sigma_0^{-1} (d\sigma/d\epsilon_{\mp}-d\sigma_B/d\epsilon_{\mp})$ to the spectrum for slow electron and positron, respectively at different values of $Z$ (continuous curves). These results are compared with the asymptotic expressions $\omega \sigma_0^{-1} d\sigma_C^{(0)}/d\epsilon_{\mp}$ (dashed curves) and $\omega \sigma_0^{-1} (d\sigma_C^{(0)}/d\epsilon_{\mp}+d\sigma_C^{(1)}/d\epsilon_{\mp})$ (dotted curves). On the one hand, one can see that for each $Z$ our results tend at large energies to the constant value $-4f(Z\alpha)$. On the other hand, the next-to-leading correction $d\sigma_C^{(1)}/d\epsilon_{\mp}$ essentially improves the agreement between exact results and asymptotic ones both for slow electron and positron.

In Ref. \cite{LMS2004} the next-to leading correction to the total cross section, $\sigma_C^{(1)}\propto m/\omega$,  was also obtained. It reads
\begin{equation}\label{cctot}
\sigma_C^{(1)}=\frac{m\sigma_0}{\omega}\left[-\frac{\pi^4}{2}\mbox{Im}\,g(Z\alpha)-4\pi(Z\alpha)^3f_1(Z\alpha)\right]\,,
\end{equation}
where the function $f_1(Z\alpha)$ is related to the total cross section $\sigma_{bf}$ of the bound-free photoproduction,
\begin{equation}\label{bf}
\sigma_{bf}=4\pi\sigma_0 (Z\alpha)^3f_1(Z\alpha)\frac{m}{\omega}\,.
\end{equation}
The function $f_1(Z\alpha)$ is of the order of unity for all values of $Z$, see Ref. \cite{LMS2004}. Since the Coulomb correction $d\sigma_C^{(1)}/d\epsilon_{\pm}$ has opposite sign for electron and positron in the region where both particles are relativistic, the correction $\sigma_C^{(1)}$ is determined by the region where momentum of electron or positron is of the order of $m$. The correction  $\sigma_C^{(1)}$ can be obtained from our results using the relation
\begin{equation}\label{sig1}
 \sigma_C^{(1)}=\int\limits_{m}^\infty d\epsilon_-\,\left[
\frac{d\sigma_C}{d\epsilon_-}+\frac{d\sigma_C}{d\epsilon_-}(Z\to -Z)+\frac{8\sigma_0}{\omega}f(Z\alpha)\right]\,.
\end{equation}
Our numerical data are in a qualitative agreement with Eq. (\ref{cctot}) though  accuracy is not very high because of strong cancellations between all terms in the integrand in Eq. (\ref{sig1}).

We conclude this section by discussing how it is possible to apply our results for photoproduction on heavy atoms, i.e., how the effects of screening can be accounted for. As it was pointed out above, the main contribution to the high-energy photoproduction cross section close to the end of spectrum of electron (or positron) comes from distances $r\sim 1/m\ll r_{\text{scr}}$, where $r_{\text{scr}}\approx 1/(mZ^{1/3}\alpha)$ is the typical screening radius. Therefore, one can account for the screening by employing the prescription formulated in Refs. \cite{Pratt1971,Pratt1972,Overbo1979}. Namely, the spectrum of slow electron in the screened Coulomb field can be obtain from that in unscreened field by means of the shift $\epsilon_-\to\epsilon_-+\Delta$ and values of the energy shift $\Delta$ for various atoms are presented in Refs. \cite{Pratt1971,Pratt1972,Overbo1979}. Analogously, for slow positron the corresponding shift is $\epsilon_+\to\epsilon_+-\Delta$. In all cases $\Delta/m < 4\times 10^{-2}$. Thus, the effect of screening in our problem is important only for very small electron or positron velocities.

\section{Conclusion}

We have calculated exactly in the parameter $Z\alpha$ the cross section of $e^+e^-$ photoproduction  in a Coulomb field at $\omega\gg m$  and $\epsilon _-\gtrsim m$ (slow electron) or $\epsilon _+\gtrsim m $ (slow positron). In the wide region, our results differ essentially from those obtained in the Born approximation as well as from the results which take into account the Coulomb corrections obtained at $\epsilon _-\gg m$ and $\epsilon _+\gg m$. Therefore, the Coulomb correction to the spectrum can be approximated by its high-energy asymptotics only at rather  large $\omega$ ($\omega \gtrsim 30\,m$).
We have found that the cross section of bremsstrahlung in a strong Coulomb field by a high-energy electron in the region where a final electron has the energy $\epsilon_1\gtrsim m$ coincides with the cross section of  $e^+e^-$ photoproduction  at $\epsilon _-\gtrsim m$ (slow electron). Finally, we have seen that the effect of screening for photoproduction close to the end of electron (positron) spectrum is important only for very small velocity of electron (positron).

\section*{Acknowledgments}
We would like to thank R. N. Lee and V. M. Strakhovenko for valuable discussions. A. I. M. gratefully acknowledges the Max-Planck-Institute for Nuclear Physics for warm hospitality and financial support during his visit. The work was supported in part by RFBR under Grant No. 09-02-00024.

\appendix
\section{Wave functions and Green's function}\label{A}
The positive-energy wave function $U_{j,l,\mu}(p,\bm r)$ with total angular momentum $j$, parity $(-1)^{j+\sigma/2}$ ($\sigma=\pm 1$), and projection $\mu$ of the total angular momentum on some quantization axis has the form \cite{BLP1980}
\begin{eqnarray}\label{wf}
&&U_{j,\sigma,\mu}(p,\bm r)=\frac{\sqrt{2}}{r}
\begin{pmatrix}
f(r)\Omega_{j,l,\mu}(\bm n)\\
-\sigma \,g(r) \Omega_{j,l',\mu}(\bm n)
\end{pmatrix}\, ,\nonumber\\
&&f(r)=\sqrt{1+\frac{m}{\epsilon}}\,\mbox{e}^{(\pi\nu/2)}\frac{|\Gamma(\gamma+1+i\nu)|}{\Gamma(2\gamma+1)}
(2pr)^\gamma\,\mbox{Im} \left\{e^{i(pr+\xi)}F(\gamma-i\nu,2\gamma+1,-2ipr)\right\}\, ,\nonumber\\
&&g(r)=\sqrt{1-\frac{m}{\epsilon}}\,\mbox{e}^{(\pi\nu/2)}\frac{|\Gamma(\gamma+1+i\nu)|}{\Gamma(2\gamma+1)}
(2pr)^\gamma\,\mbox{Re} \left\{e^{i(pr+\xi)}F(\gamma-i\nu,2\gamma+1,-2ipr)\right\}\, ,\nonumber\\
&&l=j+\frac{\sigma}{2}\,,\quad l'=j-\frac{\sigma}{2}\,,\quad \nu=\dfrac{Z\alpha\epsilon}{p}\,,\quad \kappa=\sigma \left(j+\frac{1}{2}\right)\,,\quad
\gamma=\sqrt{\kappa^2-(Z\alpha)^2}\,,\nonumber\\
&&\xi=(1-\sigma)\frac{\pi}{2}+\arctan\left[\frac{\nu(\epsilon-m)}{\epsilon (\gamma+\kappa)}\right]\,, \quad\sigma=\pm 1\,,\quad
\mbox{e}^{-2i\xi}=\frac{\kappa+i\nu m/\epsilon}{\gamma+i\nu}\,,\, \bm n=\frac{\bm r}{r}
\end{eqnarray}
where $F(\alpha,\beta_-,x)$ is the confluent hypergeometrical function and $\Omega_{j,l,\mu}(\bm n)$ is a spherical spinor. The negative-energy wave function $V_{j,\sigma,\mu}(p,\bm r)$ employed here can be obtained from $U_{j,\sigma,\mu}(p,\bm r)$ by the replacement $\epsilon\to -\epsilon$.

If $\epsilon\gg m$ and $r\sim 1/m$ then $ pr\gg 1$ and
\begin{eqnarray}\label{wfas}
&&f(r)=\sin(pr-l\pi/2+Z\alpha\ln(2pr)+\delta_\kappa)\,,\nonumber\\
&&g(r)=\cos(pr-l\pi/2+Z\alpha\ln(2pr)+\delta_\kappa)\,,\nonumber\\
&&\delta_\kappa=\xi+(l-\gamma)\frac{\pi}{2}-\mbox{arg}\,\Gamma(\gamma+1+iZ\alpha)\,,\nonumber\\
&&\xi=(1-\sigma)\frac{\pi}{4} +\arctan\left(\frac{Z\alpha}{\gamma+|\kappa|}\right)\,.
\end{eqnarray}
The high-energy asymptotics of the functions $f(r)$ and $g(r)$ for negative-energy states ($\epsilon\ll-m$) are given by Eqs.(\ref{wfas}) with the replacement $Z\to -Z$.

Let us consider the case $\epsilon\to m$, $v=p/\epsilon\ll 1$. At $|y|\to\infty$ and fixed $x$,
\begin{equation}
|\Gamma(x+iy)|\to\sqrt{2\pi}\mbox{e}^{-(\pi|y|/2)}|y|^{x-1/2}\,.
\end{equation}
Then, at $ y\to\infty$, $x\to 0$, and fixed $u=xy$ we have
\begin{equation}
F(\gamma-iy,2\gamma+1,-ix)\to \frac{\Gamma(2\gamma+1)}{u^\gamma}\left\{J_{2\gamma}(2\sqrt{u})+\frac{i}{2y}\left[
uJ_{2\gamma+2}(2\sqrt{u})-2\gamma\sqrt{u}J_{2\gamma+1}(2\sqrt{u})\right]\right\} \,,
\end{equation}
where $J_{\nu}(x)$ are ordinary Bessel functions. As a result, we find the following expressions of the low-energy asymptotic of the functions $f(r)$ and $g(r)$
\begin{eqnarray}\label{leas}
&&f(r)=\sigma\sqrt{\frac{v\pi }{Z\alpha}}\left[(\kappa-\gamma)J_{2\gamma}(2\sqrt{u})+\sqrt{u}J_{2\gamma+1}(2\sqrt{u})\right]\,,     \nonumber\\
&&g(r)=\sigma\sqrt{v\pi Z\alpha}J_{2\gamma}(2\sqrt{u})\,,\quad u=2Z\alpha mr\,.
\end{eqnarray}
It is seen that both $f(r)$ and $g(r)$ are proportional to $\sqrt{v}$ and thus are of the same order at $Z\alpha\sim 1$. However, if $Z\alpha\ll 1$ then $f(r)\gg g(r)$ at $u\sim 1$. Using the relation $J_{\nu+1}(x)=(\nu/x)J_{\nu}(x)-J_{\nu}'(x)$, where $J_{\nu}'(x)=d J_{\nu}(x)/dx$, we can write the expression for the asymptotics of the function $f(r)$ in the convenient form
\begin{eqnarray}\label{leasp}
&&f(r)=\sigma\sqrt{\frac{v\pi }{Z\alpha}}\left[\kappa J_{2\gamma}(2\sqrt{u})-\sqrt{u}J_{2\gamma}'(2\sqrt{u})\right] \,.
\end{eqnarray}

If $\epsilon\to -m$ , $v=p/|\epsilon|\ll 1$, then the functions $f(r)$ and $g(r)$ are exponentially small. We have for $Z\alpha/r\gg mv^2$:
\begin{eqnarray}\label{leas1}
&&f(r)=-\sqrt{v\pi Z\alpha}\,\mbox{e}^{-(\pi Z\alpha/v)}I_{2\gamma}(2\sqrt{u})\,,   \nonumber\\
&&g(r)=\sqrt{\frac{v\pi }{Z\alpha}}\,\mbox{e}^{-(\pi Z\alpha/v)}
\left[(\kappa+\gamma)I_{2\gamma}(2\sqrt{u})+\sqrt{u}I_{2\gamma+1}(2\sqrt{u})\right]\,,
\end{eqnarray}

In the Coulomb field, the discontinuity  of the quasiclassical electron Green's function on the cut reads \cite{MS1983a,MS1983}
\begin{eqnarray}\label{dgf}
&&\delta G(\bm r_2,\bm r_1\,|\epsilon)=\frac{ip}{4\pi}\int\limits_{-\infty}^{+\infty}\frac{d\tau}{\sinh^2\tau}
\exp\left[i2Z\alpha\dfrac{\epsilon}{p}\tau +ip(r_1+r_2)\coth\tau\right]\nonumber\\
&&\times\bigg\langle\Big[\gamma^0\epsilon+m+\dfrac{p}{2}\bm\gamma\cdot(\bm n_1-\bm n_2)\coth\tau\Big]J_0(w)
+\frac{iJ_1(w)}{w}\bigg\{\bigg[\frac{p^2(r_2-r_1)}{2\sinh^2\tau}+Z\alpha m\gamma^0\bigg]
\nonumber\\
&&\times \bm\gamma\cdot(\bm n_1+\bm n_2)  -Z\alpha p\coth\tau\,\gamma^0 [1-(\bm\gamma\cdot\bm n_2) (\bm\gamma\cdot\bm n_1)]\bigg\}\bigg\rangle\,, \nonumber\\
&& \bm n_{1,2}=\frac{\bm r_{1,2}}{r_{1,2}}\, ,\quad w=\frac{p\sqrt{2r_1r_2(1+\bm n_1\cdot\bm n_2)}}{\sinh\tau}\,.
\end{eqnarray}
In Eq. (\ref{dgf}), the contour of integration over $\tau$ passes in the positive direction around the point $\tau=0$. If $p\gg m$, $r_1\sim r_2\sim 1/m$ and $(1+\bm n_1\cdot\bm n_2)\gg m^2/p^2$, then the argument of the Bessel functions is large and Eq. (\ref{dgf}) becomes essentially simpler:
\begin{eqnarray}\label{dgf1}
\delta G(\bm x,\bm y\, ||\epsilon|)&=& -\frac{p}{2\pi R}[i\gamma^0\sin(pR+2Z\alpha s)-\bm\gamma\cdot\bm n\cos(pR+2Z\alpha s)]\,,\nonumber\\
\delta G(\bm x,\bm y\,|-|\epsilon|)&=& \frac{p}{2\pi R}[i\gamma^0\sin(pR-2Z\alpha s)+\bm\gamma\cdot\bm n\cos(pR-2Z\alpha s)]\,,\nonumber\\
\bm n&=&\frac{\bm x-\bm y}{R}\,,\quad s= \ln\left[\frac{x+y+R}{\sqrt{2(xy+\bm x\cdot\bm y)}}\right]\, ,\quad R=|\bm x-\bm y|\,.
\end{eqnarray}
\newpage

\begin{figure}[ht]
\centering
\includegraphics[height=300pt,keepaspectratio=true]{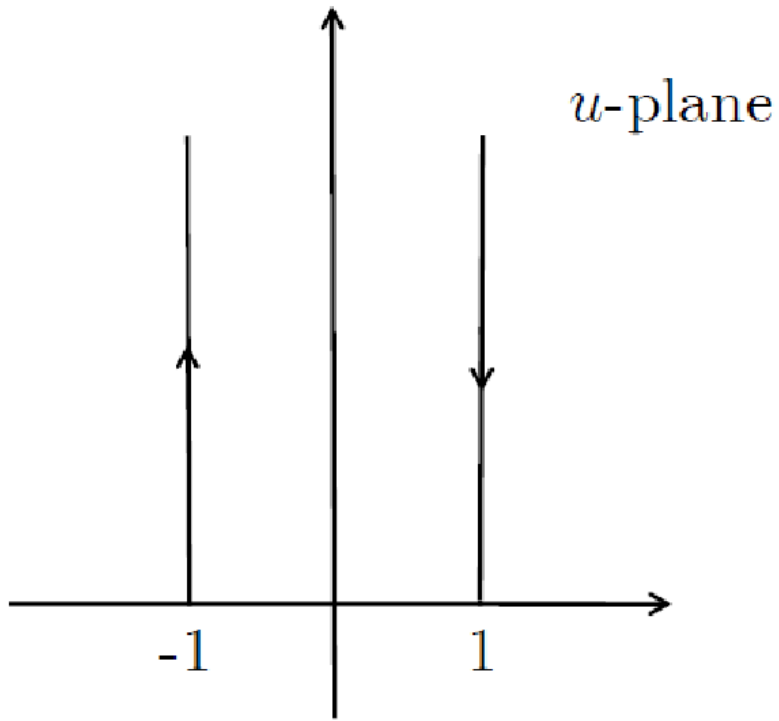}
\caption{Contour of integration in the complex plane of the variable $u$ used to perform the integral in Eq. (\ref{transform}).}
\label{CI}
\end{figure}

\newpage

\begin{figure}[ht]
\centering
\includegraphics[height=165pt,keepaspectratio=true]{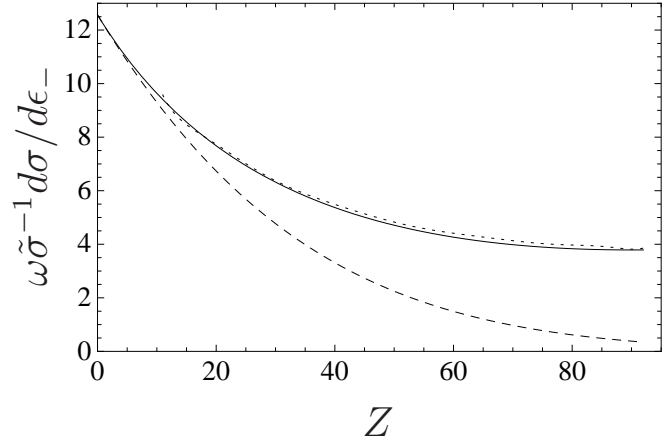}
\caption{\label{figV0} The cross section $\omega\tilde{\sigma}^{-1}d\sigma/d\epsilon_-$ of $e^+e^-$ pair production at zero electron velocity in units $\tilde{\sigma}=\alpha (Z\alpha)^3/m^2$. Solid curve: our results via Eq. (\ref{sigmaeefinalA}), dashed curve: the results of Ref. \cite{Deck1969} (see also Eq. (\ref{Deck})), dotted curve: the results of Ref. \cite{SudSharma2006} obtained at  $\omega=40\,$MeV and $\beta_-=0.1265$.}
\end{figure}

\newpage

\begin{figure}[ht]
\centering
\includegraphics[height=300pt,keepaspectratio=true]{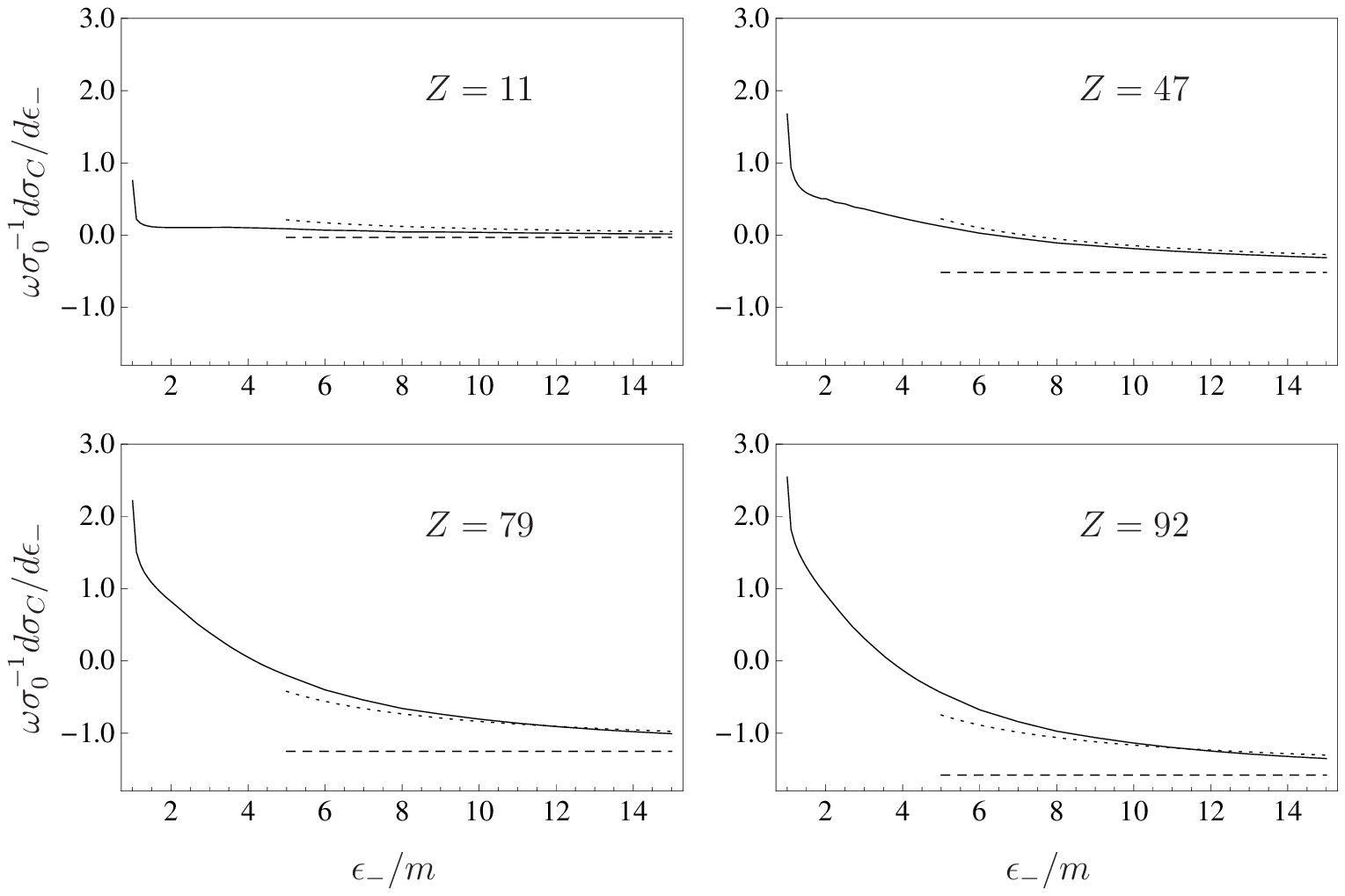}
\caption{The dependence of the Coulomb corrections $\omega\sigma_0^{-1}d\sigma_C/d\epsilon_-$ (continuous line) with $\sigma_0=\alpha(Z\alpha)^2/m^2$ on the scaled electron energy $\epsilon_-/m$ at different values of $Z$. The dashed line represents the leading-order Coulomb corrections in the limit $\epsilon_-\gg m$, while the dotted line also includes corrections proportional to $m/\epsilon_-$. The dashed and the dotted lines start at $\epsilon_-=5\,m$ because the corresponding asymptotics are valid at $\epsilon_-\gg m$.}
\end{figure}

\newpage

\begin{figure}[ht]
\centering
\includegraphics[height=300pt,keepaspectratio=true]{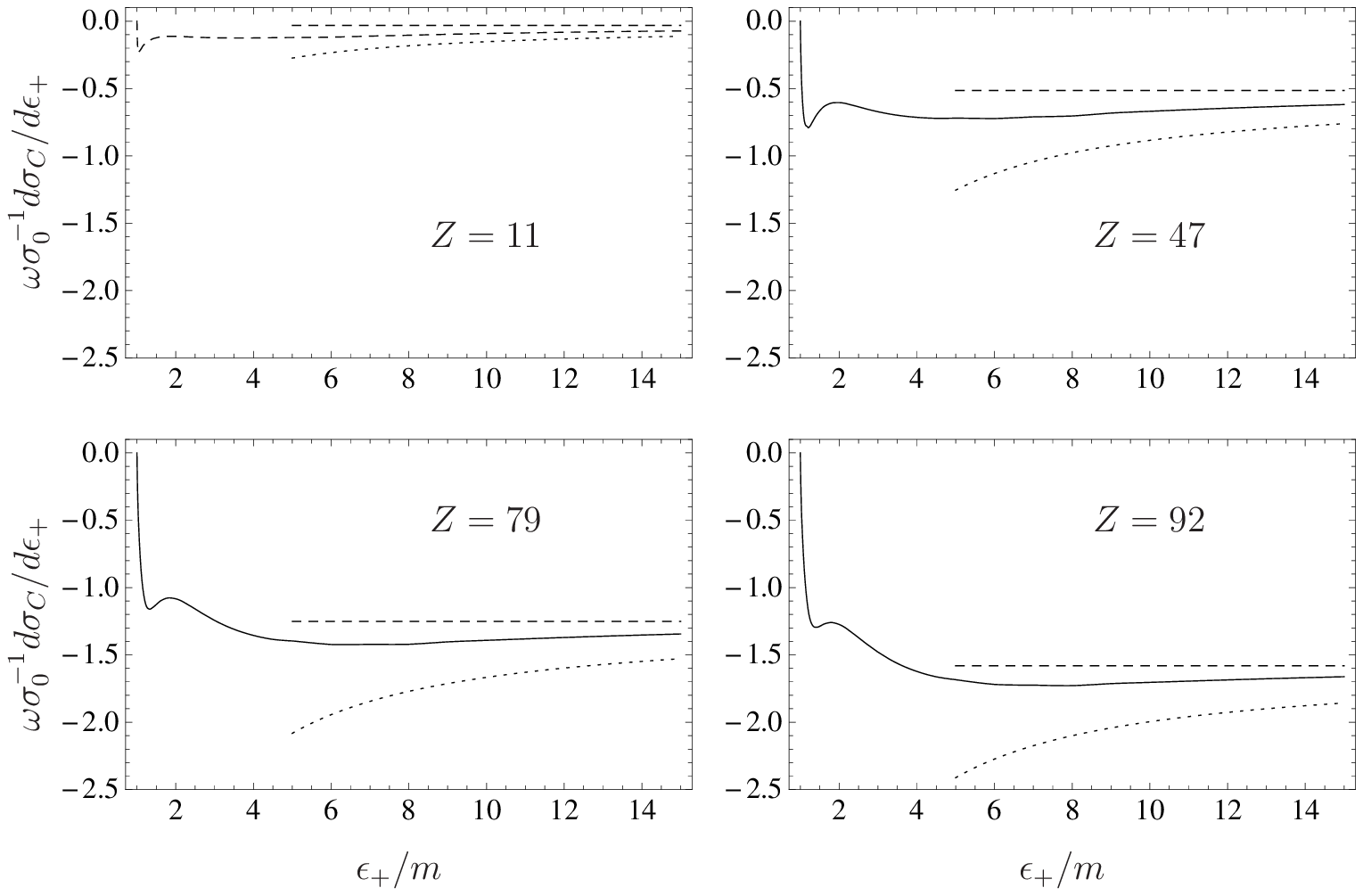}
\caption{The dependence of the Coulomb corrections $\omega\sigma_0^{-1}d\sigma_C/d\epsilon_+$ (continuous line) with $\sigma_0=\alpha(Z\alpha)^2/m^2$ on the scaled positron energy $\epsilon_+/m$ at different values of $Z$. The dashed line represents the leading-order Coulomb corrections in the limit $\epsilon_+\gg m$, while the dotted line also includes corrections proportional to $m/\epsilon_+$. The dashed and the dotted lines start at $\epsilon_+=5\,m$ because the corresponding asymptotics are valid at $\epsilon_+\gg m$.}
\end{figure}

\end{document}